\documentclass[aps,prb,twocolumn,nobibnotes,nofootinbib,superscriptaddress,showkeys,showpacs]{revtex4}
\usepackage{graphicx,amssymb,amsmath,bm}
\usepackage[english]{babel}
\DeclareMathOperator{\re}{Re}

\addto\captionsenglish{}
\addto\captionsrussian{}

\begin{document}

%\preprint{}

\title{Strong fluctuations near the frustration point in cubic lattice ferromagnets with localized moments}



\author{A.\,N.\,Ignatenko}
\email{ignatenko@imp.uran.ru}
\affiliation{Institute of Metal Physics, Kovalevskaya Str., 18, 620990, Ekaterinburg, Russia}
\author{A.\,A.\,Katanin}
\affiliation{Institute of Metal Physics, Kovalevskaya Str., 18, 620990, Ekaterinburg, Russia}
\affiliation{Ural Federal University, 620002, Ekaterinburg, Russia}
\author{V.\,Yu.\,Irkhin}
\affiliation{Institute of Metal Physics, Kovalevskaya Str., 18, 620990, Ekaterinburg, Russia}

%\author{A.\,N.\,Ignatenko\,$^{a}$, A.\,A.\,Katanin\,$^{a,b}$, V.\,Yu.\,Irkhin\,$^{a}$}
%\email[]{ignatenko@imp.uran.ru}
%\affiliation{$^{a}$Institute of Metal Physics, Kovalevskaya Str., 18, 620090, Ekaterinburg, Russia\\$^{b}$Ural Federal University, 620002, Ekaterinburg, Russia}
%\affiliation{}
%\thanks{}


\date{\today}

\begin{abstract}
Thermodynamic properties of cubic Heisenberg ferromagnets with competing exchange interactions are considered near the frustration point where the coefficient $D$ in the spin-wave spectrum $E_{\mathbf{k}}\sim D k^{2}$ vanishes. Within the Dyson-Maleev formalism it is found that at low temperatures thermal fluctuations stabilize ferromagnetism by increasing the value of $D$. For not too strong frustration this leads to an unusual ``concave'' shape of the temperature dependence of magnetization, which is in agreement with experimental data on the europium chalcogenides. The phase diagram is constructed by means of Monte Carlo simulation, and suppression of magnetization and Curie temperature is found in comparison with the results of the spin-wave theory. This effect is explained by the the presence of non-analytical corrections to the spin-wave spectrum which are represented in the lowest order by the term $\sim (T/S)^{2} k^{2}\log{k}$.
\end{abstract}

%\pacs{75.10.Jm, 75.30.Ds, 64.60.De, 75.10.Hk, 75.40.Mg, 75.40.-s, 75.30.Kz}
%75.10.Jm 	Quantized spin models, including quantum spin frustration 
%75.30.Ds 	Spin waves (for spin-wave resonance, see 76.50.+g)
%64.60.De 	Statistical mechanics of model systems (Ising model, Potts model, field-theory models, Monte Carlo techniques, etc.) 
%75.10.Hk 	Classical spin models
%75.40.Mg 	Numerical simulation studies
%75.40.-s   	Critical-point effects, specific heats, short-range order (for equilibrium properties near critical points, see 64.60.F-; for dynamical critical phenomena, see 64.60.Ht)
%75.30.Kz 	Magnetic phase boundaries (including classical and quantum magnetic transitions, metamagnetism, etc.) (for ferroelectric phase transitions, see 77.80.B-; for superconductivity phase diagrams, see 74.25.Dw)


\maketitle


%   -------------------------------------------------------------------------------------------------------------------------------------------------------------------------------
%   ----------------- Begin   Main Text ---------------------------------------------------------------------------------------------------------------------------------------
%   -------------------------------------------------------------------------------------------------------------------------------------------------------------------------------

	During last thirty years, the physics of quantum magnets has been developed under strong influence of the concept of quantum disordered ground state, which gives an alternative to the standard magnetic order picture. Main results in this direction were obtained for antiferromagnets in one and two dimensions (see, e.g., Ref. \onlinecite{MagBook}). However, recently the existence of a quantum disordered ground state in \emph{two-dimensional} magnets with predominantly \emph{ferromagnetic} interactions was demonstrated, see Refs. \onlinecite{ShannonPRL06, Zhitomirsky_2010}. The formation of the corresponding quantum state, called a spin nematic, is associated with an instability with respect to the creation of bound states of spin waves. Since in the ferromagnetic ground state quantum fluctuations are completely absent, the physics of the ``frustrated ferromagnets'' differs substantially from the physics of antiferromagnets with competing exchange interactions.    

	Magnetic frustrations also significantly influence the thermodynamic properties of three-dimensional systems, in particular the temperature dependence of long- and short-range order parameters (see e.g. Ref. \onlinecite{JETPl_2008} for the fcc antiferromagnet). Anomalous temperature dependences of the magnetization are indeed observed experimentally in ferromagnetic materials, for example in the overdoped europium chalcogenides, see Ref. \onlinecite{EuO_PRL2010}. Typically, these anomalies are explained by introducing into the Hamiltonian four-spin (biquadratic) interactions \cite{Nagaev}. In this paper we show that these anomalies can be naturally obtained in the framework of the standard Heisenberg model by taking into account the frustration consistently.

	We consider the Heisenberg model to investigate thermal properties of three-dimensional frustrated ferromagnets. The frustration is caused by the exchange interactions in higher coordination spheres. Low-temperature properties of this model can be determined not only by the conventional magnons, but also by the bound states of spin waves \cite{Dyson_1956, Wortis_1963}. However, as it was shown in Ref. \onlinecite{Rastelli_BoundSW1991}, in \emph{three} dimensions frustration does not lead to the formation of the bound state of two spin waves at long wavelengths. Therefore, the contribution of the bound states is absent at low temperatures.

	The spin-wave spectrum $E_{0}(\mathbf{k})=S\left[J(0)-J(\mathbf{k})\right]$ (known exactly for a ferromagnet at zero temperature) is quadratic in the wave vector at $\mathbf{k}\to 0$, $E_{0}(\mathbf{k})\approx D_{\alpha\beta} k_{\alpha} k_{\beta}$, where $D_{\alpha\beta}=S\sum_{\mathbf{r}} J_{r} r_{\alpha}r_{\beta}/6$ is the spin-wave stiffness tensor. Here $J(\mathbf{k})$ is the Fourier transform of the exchange interaction $J_{r}$, $r_{\alpha}$ are the coordinates of the lattice sites. Provided that owing to the frustration two or three eigenvalues of $D_{\alpha\beta}$ vanish, in the equation for the magnetization in the spin-wave theory 
\begin{equation}  
\label{mag_eq} 
M_{\text{SWT}} = S-\int\frac{d^{3} \mathbf{k}}{(2\pi)^3} \frac{1}{\exp{[E_{0}(\mathbf{k})/T]}-1}
\end{equation}
the second term diverges at any finite temperature $T$. The exact form of the divergence is determined by the higher order terms of the expansion of $E_{0}(\mathbf{k})$ in $\mathbf{k}$. Obviously, when all three eigenvalues of $D_{\alpha\beta}$ vanish simultaneously, the divergence is stronger. This case is easily realized for cubic crystals where the tensor $D_{\alpha\beta}=D_{0}\,\delta_{\alpha\beta}$ is always diagonal. Below we assume the presence of cubic symmetry.

	Corrections to the spin-wave theory can be calculated systematically using the Dyson-Maleev representation for spin operators \cite{Dyson_1956}. The corresponding effective boson Hamiltonian is $H_{\text{eff}}=H_{0}+H_{\text{int}}$ where
\begin{equation}
H_{0}=\sum_{\mathbf{k}} E_{0}(\mathbf{k}) a^{\dagger}_{\mathbf{k}}a^{}_{\mathbf{k}}
\end{equation} 
is the Hamiltonian of non-interacting magnons,
\begin{equation}
H_{\text{int}}=\frac{1}{4N}\sum_{\mathbf{k},\mathbf{k}',\mathbf{q},\mathbf{q}'} \varphi\left(\mathbf{k},\mathbf{k}';\mathbf{q},\mathbf{q}' \right) a^{\dagger}_{\mathbf{k}}a^{\dagger}_{\mathbf{k}'}a^{}_{\mathbf{q}}a^{}_{\mathbf{q}'}\delta_{\mathbf{k}+\mathbf{k}',\mathbf{q}+\mathbf{q}'}, %\left(1+2-3-4\right)
\end{equation}
is the interaction, $\varphi\left(\mathbf{k},\mathbf{k}';\mathbf{q},\mathbf{q}' \right)= J(\mathbf{q})+J(\mathbf{q}')-J(\mathbf{k}-\mathbf{q})-J(\mathbf{k}-\mathbf{q}')$, see Ref. \onlinecite{BKJbook}. The spin-wave spectrum at finite temperatures, $E=E_{0}+\re\Sigma$, is determined by the self-energy of bosons $\Sigma(\mathbf{k},i \omega_{n})$. Fig. 1 shows low-order Feynman diagrams for $\Sigma$. 
\begin{figure}[hpt]
%FIG. 1
\includegraphics[width=5cm]{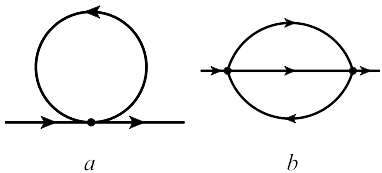}
\caption{Diagrams for boson self-energy in the first and second orders in $1/S$.}
\end{figure}

	Let us consider the contribution of the diagram (a) to the spin-wave stiffness at finite temperature, $D(T)$. Direct evaluation shows that at low $T\ll D_{0}$
\begin{equation}
\label{rho_lowT}
D=D_{0}-A \left(T/D_{0}\right)^{5/2}.
\end{equation} 
Note that the coefficient $A$ can be expressed in terms of the coefficients of the expansion of  $E_ {0}(\mathbf {k})$ in $ \mathbf{k}$. For cubic crystals there are only two fourth-order invariants,
\begin{equation}
\label{spectrum_expansion}
E_{0}(\mathbf{k})\approx D_{0}k^{2}+\varkappa_{1}k^{4}+\varkappa_{2}(k_{x}^{4}+k_{y}^{4}+k_{z}^{4}).
\end{equation}
Then $A=-\zeta(5/2)(5\varkappa_{1}+3\varkappa_{2})/8\pi^{3/2}S$. Under normal conditions, i.e. in the absence of the frustration, the combination $5\varkappa_{1}+3\varkappa_{2}=-S\sum_{\mathbf{r}}J_{r}r^{4}/6<0$.  Therefore the coefficient $A>0$ and with increasing temperature the spin-wave stiffness decreases and the magnetic order is destroyed. However, if the frustration is present, the situation may change \cite {Rastelli_BoundSW1991, Heinila_PRB1993}.  We assume below that even for an arbitrarily small $D_{0}$ the ferromagnetic \emph{ground state} remains stable. The requirement of the positivity of Eq. (\ref{spectrum_expansion}) for $D_{0}\to 0$ then gives inequalities $\varkappa_{1}+\varkappa_{2}>0$, $3\varkappa_{1}+\varkappa_{2}> 0$. One can see that in the considering case these inequalities lead to $A<0$. For $D_{0}=0$ we find a behavior, analogous to Eq. (\ref{rho_lowT}), but with a different exponent,
\begin{equation}
\label{rho_highT}
D(T)=- b A \left(T/\varkappa\right)^{5/4}
\end{equation}
where $T \ll \varkappa$, $\varkappa=\text{max}\{\varkappa_{1},\varkappa_{2}\}$, $b>0$ is the function of $\varkappa_ {1,2}$. Rearranging the diagram series in the occupation numbers of bosons \cite{BaymSessler_1963}, one can show that the dependences (\ref{rho_lowT}) and (\ref {rho_highT}) do not change their form in higher orders in $1/S$. Thus, regardless of the details of the exchange interactions, in the frustration regime the ferromagnetic order at low temperatures is stabilized by thermal fluctuations (similarly to the ``order from disorder'' effect, see, e.g., Ref. \onlinecite{Villain1980}).

	The nonlinear spin-wave theory (which takes into account only the diagram Fig. 1a) is valid in the limit $T\to 0$. We have furthermore performed the calculations within the self-consistent spin-wave theory (SSWT), previously used for two- and three-dimensional systems \cite{IKK_PRB_USP}. SSWT includes all kinds of diagrams obtained from the diagram (a) by its recursive reinserting into internal lines. To improve the behavior of SSWT results at high temperatures, the contributions from pseudofermions are included \cite{BKJbook}, the chemical potential of bosons being introduced to fulfil the equation $M=0$ in the paramagnetic phase \cite{TakahashiPRB1989}. Fig. 2 shows an example of dependences $D(T)$ calculated in SSWT for $S=1/2$ Heisenberg model on the fcc cubic lattice. It can be seen that at low temperatures near the frustration point $D_{0}=0$ the dependences $D(T)$ are consistent with Eqs. (\ref{rho_lowT}) and (\ref{rho_highT}).
\begin{figure}[hpt]
%FIG. 2
\includegraphics[width=8cm]{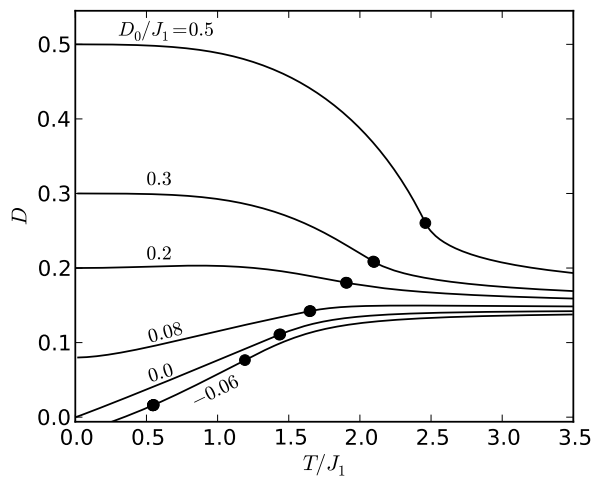}
\caption{Temperature dependence of the spin-wave stiffness calculated in SSWT for $S=1/2$ Heisenberg model on the fcc cubic lattice with the interaction between the first, second, and third neighbors. $J_{2}=-0.3 J_{1}$, and $J_{3}=\left(D_{0}/S-J_{1}-J_{2}\right)/6$ was chosen for a number of values of $D_{0}$ shown in the figure. The filled circles correspond to the Curie temperatures.}
\end{figure}

	Note that the spectrum $E(\mathbf {k})$ in SSWT equation for the magnetization, Eq. (\ref{mag_eq}), is renormalized by temperature corrections. As a result, the magnetization does not diverge at $D_{0}=0$. However, the growth of the spin-wave stiffness with increasing temperature, see Eq. (\ref{rho_highT}), results in this case in unusual behavior of the magnetization, $M\simeq S- uT^{3/8}$. Fig. 3 shows the temperature dependence of the magnetization for several values of $D_{0}$. It can be seen that for small non-negative $D_{0}$ the curves $M(T)$ at low temperatures acquire a concave form. Such anomalous dependences were observed experimentally in overdoped EuO \cite {EuO_PRL2010}.
\begin{figure}[hpt]
%FIG. 3
\includegraphics[width=8cm]{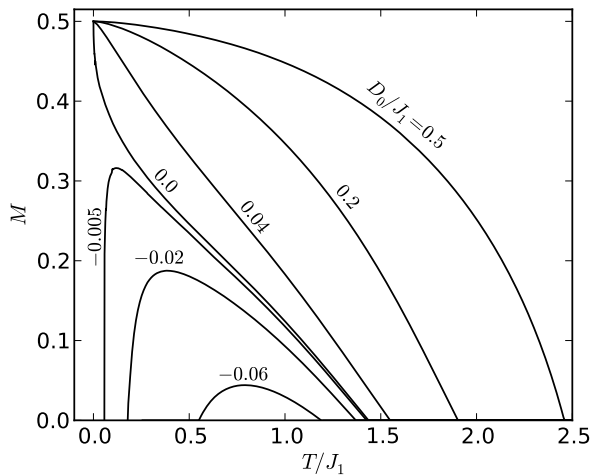}
\caption{The dependence of magnetization on temperature in SSWT. The model parameters are chosen as described in the caption to Fig. 2.}
\end{figure}

	The results for spin-wave stiffness and magnetization at negative $D_{0}$ (when the ferromagnetic ground state becomes unstable) are also shown in Figs. 2 and 3. In this case the temperature corrections lead to the stabilization of the ferromagnetic ordered phase above some ``lower'' critical temperature. Therefore, the dependence of the Curie temperature $T_{c}$ on $D_{0}$ becomes two-valued at $D_{0}<0$ (see Fig. 4 for a classical system). Note that classical systems can be studied in the same way as the quantum systems (for this it is necessary to replace $J_{ij}\to J_{ij}/S^{2}$ in all equations and then pass to the limit $S\to\infty$). All the results obtained in the quantum case do not change qualitatively in the classical limit (quantitative differences exist, in particular temperature exponents in Eqs.  (\ref{rho_lowT}) and (\ref{rho_highT}) will change). This is due to the weakness of the quantum effects in the ferromagnetic state. Below we will use this fact to identify the limits of applicability of SSWT by comparing SSWT with the results of computer simulation for classical systems.

	When $D<0$ and ferromagnetism is unstable, another state, which is able to compete with the ferromagnetic state, should appear. In principle, this could be a spin liquid state. However, the most likely candidate is a spiral state. Note that for $D<0$ the spin-wave spectrum (\ref{spectrum_expansion}) has a minimum at $\mathbf{k}=\mathbf{Q}_{+}=(Q_{+},Q_{+},Q_{+})$, $Q_{+}=\left(-D/2[3\varkappa_{1}+\varkappa_{2}]\right)^{1/2}$, in the case $\varkappa_{2}>0$, and at $\mathbf{k}=\mathbf{Q}_{-}=(Q_{-},0,0)$, $Q_{-}=\left(-D/2[\varkappa_{1}+\varkappa_{2}]\right)^{1/2}$, in the case $\varkappa_{2}<0$. This could mean the formation of a spiral state with the wave vector $\mathbf {Q}_{+}$ or $\mathbf {Q}_{-}$. This state appears to be favourable in particular in the mean field approximation. Note that the ferromagnetic and paramagnetic states predicted in SSWT without account for spirals (which account is difficult) can be metastable. It is important that the continuous transition from ferromagnetic to spiral state is prohibited in SSWT because this transition would require $D(T)$ to vanish at the boundary between two phases. The latter is impossible due to the divergence in the magnetization, Eq. (\ref{mag_eq}), for $D(T)\to 0$ and $T>0$. 

	For a complete study of the phase diagram and determination of the role of non-spin-wave contributions, we have performed Monte Carlo simulation of the classical Heisenberg model on the finite fcc cubic lattices of extent $L=10, 14$ in each direction with the help of the ALPS library \cite{ALPS2_0}. The resulting phase diagram is also shown in Fig. 4. It contains ferromagnetic (F), spiral (S), and paramagnetic (P) phases. The point of frustration $D_{0}=D_{\text{F}}(L)$, which defines the boundary between the ferromagnetic and the spiral state at $T=0$, is slightly shifted to negative $D_{0}$ due to the discreteness of the quasimomentum of finite systems. We have found that, for $D_{0}<D_{\text{F}}$ and at temperatures below the ferromagnetic phase, the paramagnetic phase does not arise. Instead, the first-order transition to the spiral phase occurs (we have observed a two-peak structure of the probability distribution for the magnetization). This means that for a given finite system the curve $T_{c}(D_{0},L)$ near $T=0$ lies in the region of metastable states, i.e. below and to the left from the line of first-order phase transitions. At the same time, in the quantum case the curve $T_{c}(D_{0})$ near $T=0$ may correspond to the transition to the quantum disordered phase.
\begin{figure}[hpt]
%FIG. 4
\includegraphics[width=8cm]{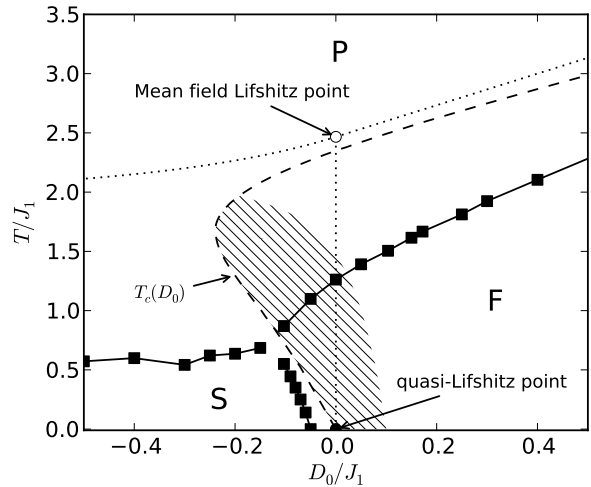}
\caption{Phase diagram of the classical Heisenberg model on the fcc cubic lattice with the interaction of the first, second, and third neighbors, $J_{2}=-0.3 J_{1}$, $J_{3}=\left(D_{0}-J_{1}-J_{2}\right)/6$. Solid line is the result of Monte Carlo simulation for $10\times 10\times 10$ system.  Dotted line is the mean field approximation. Dashed line shows the dependence of the Curie temperature on $D_{0}$ in SSWT. The shaded area is the part of the ferromagnetic phase with $D(T)<0.1 J_{1}$ according to SSWT.}
\end{figure}

	For comparison, Fig. 4 also shows the phase boundaries in the mean field approximation. A peculiarity of this approximation is the existence of the multicritical Lifshitz point \cite{Hornreich_PRL1975} which is the endpoint of the boundary between ferromagnetic and paramagnetic phases. In the Ginzburg-Landau expansion at this point the coefficient $c\propto D(T)$ at $[\nabla M(x)]^{2}$ vanishes. This would lead to a modification in the critical behavior \cite{Hornreich_PRL1975}. However, taking into account fluctuation corrections to the mean field theory in SSWT makes impossible the existence of the Lifshitz point at finite temperatures due to the divergence in Eq. (\ref{mag_eq}) for $D(T)\to 0$. On the curve $T_{c}(D_{0})$ describing the boundary between the ferromagnetic and paramagnetic phases in SSWT, the coefficient $D(T)$ is finite everywhere and vanishes only in the limit $T \to 0$. In this sense, one can say that fluctuation corrections shift the Lifshitz point along the curve $T_{c}(D_{0})$ to the point with coordinates $T=+0$, $D_{0}=-0$. Infinitely small deviation of $T$ and $D_{0}$ from zero is essential, since strictly at $T=0$ even in the quantum case the system possesses saturated ferromagnetism for all $D_{0}\ge 0$. Thereby we call the point with coordinates $T=+0$, $D_{0}=-0$ the \emph{quasi-Lifshitz point}. In SSWT, when moving away from the quasi-Lifshitz point along the curve $T_{c}(D_{0})$, the size of the corresponding ``critical'' temperature range (where the magnetization changes from its maximum value to zero) increases as $(-D_{0})^{4/5}$ (see also Fig. 3). Therefore, given that the curve $T_{c}(D_{0})$ in the neighborhood of $T=0$ falls into the region of metastable states, fluctuations in the region of stable states are expected to be important at sufficient distance from the quasi-Lifshitz point near the curve $T_{c}(D_{0})$. The quasi-Lifshitz point affects also the behavior of the magnetization for weakly positive $D_{0}$, see Fig. 3.
\begin{figure}[hpt]
%FIG. 5
\includegraphics[width=8cm]{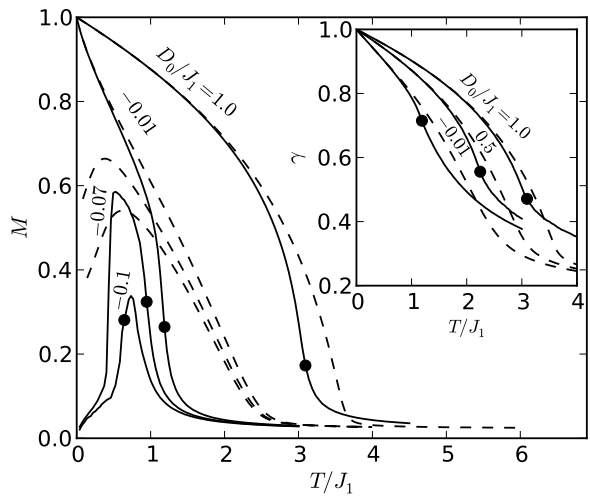}
\caption{Temperature dependence of the magnetization for the classical Heisenberg model on the fcc cubic lattice of finite size $14\times 14\times 14$. The model parameters are chosen as in the caption to Fig. 4. The inset shows temperature dependences of the short-range order parameter. Solid lines are the results of Monte Carlo simulation. Dashed lines are the result of SSWT. The filled circles indicate the Curie temperatures determined by the maximums of the susceptibility.}
\end{figure}

	Everywhere, except for sufficiently large $D_{0}\approx J_{1}$, the Curie temperature in SSWT and mean-field approximation is overestimated by 2-3 times in comparison with the results of the numerical simulations, see Fig. 4 (cf. Ref. \onlinecite{Heinila_PRB1993}). Fig. 5 shows that for $D_{0}\approx J_{1}$ SSWT gives a good description of the temperature dependence of the magnetization even at high temperatures, except for a narrow critical region near $T_{c}$. In the presence of frustration for $D_{\text{F}}(L)<D_{0}\ll J_{1}$ the ferromagnetic order in the ground state is stable, and SSWT still reliable at low temperatures, but becomes inapplicable at higher temperatures. The latter fact is manifested, in particular, in a significant overestimation of the Curie temperature. In the presence of strong frustration for $D_{0}<D_{\text{F}}(L)$ ferromagnetism in Monte Carlo simulation is observed only at intermediate temperatures, which is in a qualitative agreement with SSWT, but the quantitative discrepancies appear now at all temperatures.

	The inset in Fig. 5 shows the temperature dependence of the short-range order parameter $\gamma=\sqrt{\mathbf{S}_{i}\cdot\mathbf{S}_{j}}$ ($i$ and $j$ are nearest neighbors). In the presence of frustration the short range order persists up to $T_{c}$ and above, as one would expect. It is remarkable that even in the case of strong frustration and high temperatures SSWT describes the short-range order much better than the long-range one. This means that the basic defect of SSWT for strong frustration is associated with insufficient account for long-wavelength fluctuations. The existence of short-range order can be crucial for the energy gain of certain phases, which, in particular, is of interest for the description of the $\gamma$-$\alpha$ transition in iron \cite{JETPl_2008, Gornostyrev_2009}.

	Note that when the system is frustrated, i.e. for $D_{0}\ll J_{1}$, according to the results of Fig. 2 the spin-wave stiffness $D(T)$ is small even at temperatures up to the Curie temperature and above. This can explain the failure of SSWT in the case of strong frustration. If we formally consider $D(T)$ as an independent parameter, a perturbation series in $1/S$ for various physical quantities will contain divergent terms in the limit $D(T)\to 0$. An example is given by the Eq. (\ref{mag_eq}) for the magnetization. The singularities appear also in the expansion of the self-energy. Consider, e.g., two-loop diagram (Fig. 1b). In the limit $D(T)\to 0$ it gives singular term in the spin-wave spectrum
\begin{equation}
\label{log_term}
E(\sqrt{D/\varkappa_{1}}\ll k \ll \Lambda) \sim \varkappa_{1} k^{4}+\frac{T^{2}}{6\pi^{2} S^{2}\varkappa_{1}}k^{2}\log\frac{k}{\Lambda}
\end{equation}
($\varkappa_{2}=0$, $\Lambda$ is an ultraviolet cutoff; the condition for the applicability of the perturbation theory gives an additional estimate for the momentum, $k^{2}/\log{\left(\Lambda/k\right)} \gg 0.02 (T/S\varkappa_{1})^{2}$). Note that the logarithmic term is negative. Similar logarithmic terms are typical for the systems with strong fluctuations (see, e.g., Refs. \onlinecite{Kosevich_JETP_1986}, \onlinecite{IKK_PRB_USP}).

	It is important that the singular terms similar to that in Eq. (\ref{log_term}) can result in a significant suppression of the magnetic order even at finite but small $D(T)$. Suppression of the magnetic order should be  particularly strong near the lower critical temperature (see the turn of the curve $T_{c}(D_{0})$ in Fig. 4) since the value of $D(T)$ is smaller in this case (see Fig. 2).

	As it was found above, $D(T)$ does not vanish at finite temperatures due to the divergence in Eq. (\ref {mag_eq}). However, because of the presence of singular corrections to the spectrum (see Eq. (\ref{log_term})), the scenario in which the spin-wave spectrum acquires non-analytic form $E(\mathbf{k}\to 0)\sim k^{a}$, $2<a<3$, for $D=0$ is in principle not excluded \footnote{We thank the referee for drawing our attention to this possibility.}. This spectrum does not lead to a divergence in the magnetization, which allows $D$ to vanish. In this case a continuous transition between the ferromagnetic and helical phases, as well as the existence of the Lifshitz point at finite temperature, are possible. However, in our numerical simulations we did not find any confirmation of this scenario (also, one can show that this scenario does not appear in the spherical model and $4-\varepsilon$ expansion).

	The results, similar to those discussed in the paper, can be obtained for all cubic lattices (simple cubic, bcc, and fcc) with different types of competing interactions. The presence of low-temperature regions with strong fluctuations on the phase diagram could favor the formation of a quantum disordered state in three-dimensional frustrated ferromagnets. To achieve advance in this direction, taking into account quantum effects in the spiral state near the boundary with ferromagnetic phase is of considerable interest. The ``bold'' diagrammatic Monte Carlo technique developed quite recently in Ref. \onlinecite{BoldDiagMC} can provide the appropriate tool for this.

	We are grateful to Yu.N.~Gornostyrev for valuable discussions. This work is partly supported by ``Dynasty'' foundation, the project of young scientists of Ural Branch of RAS, project No. M-9, and the Programs of fundamental research of RAS Physical Division ``Strongly correlated electrons in solids and structures'', project No. 12-T-2-1001 (Ural Branch) and of RAS Presidium ``Quantum mesoscopic and disordered structures'', project No. 12-P-2-1041. Monte Carlo simulation was performed using ``Uran'' cluster of IMM UB RAS.

\end{document}